\documentclass[preprints,article,submit,moreauthors,dvi2pdf]{mdpi}

\firstpage{1} 
\pubvolume{xx}
\issuenum{1}
\articlenumber{5}
\pubyear{2019}
\copyrightyear{2019}
\makeatletter
\@ifundefined{linenumbers}{}{%
	\let\linenumbers\relax
	
}
\makeatother

\history{Received: date; Accepted: date; Published: date}

\Title{$\beta^-$-decay half-lives of even-even nuclei using
  the recently introduced phase space recipe}

\Author{Jameel-Un Nabi $^{1}$, Mavra  Ishfaq $^{1,2}$, Ovidiu Ni\c{t}escu $^{3,4,5}$ , Mihail Mirea $^{4,5}$, Sabin Stoica $^{4,5}$}

\AuthorNames{Firstname Lastname, Firstname Lastname and Firstname Lastname}

\address{%
$^{1}$ \quad Affiliation $^{1}$GIK Institute of Engineering Sciences and Technology,Topi 23640, Khyber Pakhtunkhwa, Pakistan; \\
$^{2}$ \quad Affiliation $^{2}$ The University of Lahore, Gujrat, Punjab, Pakistan; \\
$^{3}$ \quad Affiliation $^{3}$ University of Bucharest, Faculty of Physics, P.O. Box MG11, 077125-Magurele, Romania\\
$^{4}$ \quad Affiliation $^{4}$ National Institute of Physics and Nuclear Engineering, P.O. Box MG6, 077125-Magurele, Romania\\
$^{5}$ \quad Affiliation $^{5}$ International Centre for Advanced Training and Research in Physics, P.O. Box MG12, 077125-Magurele, Romania\\}

\corres{Corresponding author email: jameel@giki.edu.pk;}

\abstract{We present the $\beta$-decay half-lives calculation for selected even-even nuclei that decay through electron emission. The kinematical portion of the half-life calculation was performed using a recently introduced technique for computation of phase space factors (PSFs). The dynamical portion of our calculation was performed within the proton-neutron quasiparticle random phase approximation (pn-QRPA) model. Six nuclei ($^{20}$O, $^{24}$Ne, $^{34}$Si, $^{54}$Ti, $^{62}$Fe and $^{98}$Zr) were selected for the present calculation. We compare the calculated PSFs for these cases against the traditionally used recipe. In our new approach, the Dirac equation was numerically solved employing a Coulomb potential. This potential was adopted from a more realistic proton distribution of the daughter nucleus. Thus, the finite size of the nucleus and the diffuse nuclear surface corrections are taken into account. Moreover, a screened Coulomb potential was constructed to account for the effect of atomic screening.  The power series technique was used for the numerical solution. The calculated values of half-lives, employing the recently developed method for computation of PSFs, were in good agreement with the experimental data.  
}


\begin{document}

\section{Introduction}
In the last decades, the $\beta$-decay process shaped our perspective of modern physics. From changing the evolution of the Standard Model and revealing the nature of left-handed '$V-A$' weak
interaction \cite{Wei09}, theoretical study of $\beta$-decay, also,
performed a key part in the understanding of astrophysical processes
e.g. nucleosynthesis ($r-$, $rp-$, $s-$, $p-$) processes and presupernova evolution of massive stars \cite{Mol03,Ni14,Ren14}. Probing different observables of $\beta$-decay, continue to be at the forefront of new physics searches, but dealing with the Hamiltonian
that administers all types of $\beta$-decay is not a trivial problem and
involves approximations.\-

The weak interaction which produces the real decay is much weaker as compared to the electromagnetic interaction of beta-particle with its neighborhood. Consequently, the later interaction may not be treated in the perturbation theory for nuclear $\beta$-decay Hamiltonian \cite{Gro76, Hal70}. The typical approximation is to use the Dirac equation having an electrostatic potential instead of a plane-wave of $\beta$-particle \cite{Rom65}. With this replacement, the half-life for nuclear $\beta$-decay is calculated after the computation of associated nuclear matrix elements (NMEs) and PSFs.

In literature various approaches of $\beta$ spectrum description and PSFs calculation were reported \cite{Beh68,Beh70,Kon41,Hay18}. Some of these calculations employed the point charge Fermi function \cite{Fer34}. A realistic PSF must evaluate the $\beta$-particle radial wave function as solutions to the Dirac equation in a finite charge distribution that adequately describe that of the real nucleus \cite{Wil90}. In this paper, we used $\beta$-particle exact radial
wave functions for the construction of the Fermi function acquired by solving the Dirac equation numerically with realistic electrostatic potential. In this approach, we included electrostatic finite-size corrections (finite size and diffuse nuclear surface) and atomic screening corrections by constructing appropriate Coulomb potential. The numerical recipe used to solve the Dirac equation limited the truncation errors of solutions, and the only remaining discrepancies were because of rounding errors and distortion of potential
initiated by interpolating spline \cite{Sal91, Sal95}. The
present technique for the calculation of PSFs could
easily be developed for any arbitrary nucleus. For further details
about the new recipe of PSFs, we refer to \cite{Sab16}.

Another consequence of the approximation discussed above is that the
PSF calculation and $\beta$ spectrum predictions must include the electromagnetic correction. Examples for such correction are the emission of internal bremsstrahlung during decay and the interaction of $\beta$-particle with decaying nucleon. Hayen \textit{et al.}\cite{Hay18} tabulated all corrections that should be applied on $\beta$ spectrum calculation. This type of correction is not included in the recently introduced recipe for calculation of PSF \cite{Sab16}.

For the current half-lives calculation of $\beta$-decay, NMEs were calculated within the framework of (pn-QRPA) model in a deformed basis and a schematic separable potential, both in
particle-hole and particle-particle channels. In this work we
compute PSFs and $\beta^{-}$-decay half-lives for
six even-even nuclei ($^{20}$O, $^{24}$Ne, $^{34}$Si, $^{54}$Ti,
$^{62}$Fe and $^{98}$Zr). We compare our PSFs \cite{Sab16} with the ones computed by Gove and Martin (GM) \cite{Gov71}. We later calculate the half-lives of $\beta$-decay for these selected nuclei using the traditional \cite{Gov71} and newly introduced \cite{Sab16} recipes for computation of PSFs. The half-lives are further compared with the measured values.

We briefly describe the necessary formalism of reported work in
Section~2. We discuss our results and compare them with measured
data in Section~3. We report our findings in Section~4.

\section{Theoretical Framework}

The theoretical framework for computation of PSFs for any
type of nuclear $\beta$ transition is  presented in detail in Ref. \cite{Gov71}. We included only allowed transitions (Fermi
and Gamow-Teller type) in our calculations. The traditional PSF for an allowed $\beta$
transition is given in natural units ($\hbar = m = c = 1$) by
\begin{equation}\label{ps1}
f=\int_1^{E_0} pE(E_0-E)^2F(Z,E)dE.
\end{equation}

In "Eq.~(\ref{ps1})" $p$ denotes $\beta$-particle momentum,
$E$ = $\sqrt{p^2 + 1}$ stands for total energy while $E_0$ is the
maximum energy of $\beta$-particle. Many authors \cite{Kon41,Beh69} start from the point charge
Fermi function \textbf{$F(Z,E)$};

\begin{equation}\label{eq:FermiFunc}
F(Z,E) =4(2pR)^{2(\gamma-1)}e^{\pi\eta}\frac{\left|\Gamma(\gamma+i\eta)\right|^{2}}{[\Gamma(2\gamma+1)]^{2}}
\end{equation}
introduced in 1934 \cite{Fer34}. In "Eq.~(\ref{eq:FermiFunc})"
$\alpha\simeq1/137$ is the fine structure constant; $\gamma =
\sqrt{1-\alpha^{2}Z^{2}}$; $\eta=\alpha Z E/p$; $R$ is the
radius of the nucleus in units of $\hbar/ m_e c^2$; and
$\Gamma$ is the gamma-function. Fermi function could be described in terms of the radial wave functions
\begin{equation}\label{eq:norm}
F(Z,E,r) = {f^2_{1}(Z,E,r)+g^2_{-1}(Z,E,r)\over 2p^2},
\end{equation}
which, historically, has been evaluated at the nuclear radius
or the origin. In this paper we consider the evaluation at the
nuclear surface. The functions $f_1(Z,E,r)$ and $g_{-1}(Z,E,r)$ are
small and large radial wave functions, respectively, and satisfy
coupled equations \cite{Ros61},
\begin{eqnarray}\label{dirac}
\begin{array}{rcl}
\left(\frac{d}{dr} +
\frac{\kappa+1}{r}\right)g_{\kappa}(E,r) &=&
(E+V(r)+1)f_{\kappa}(E,r) \\[8pt]
\left(\frac{d}{dr} + \frac{\kappa-1}{r}\right)f_{\kappa}(E,r)&=&
-(E+V(r)-1)g_{\kappa}(E,r),
\end{array}
\end{eqnarray}
where $V(r)$ is the central potential of $\beta$-particle while
$\kappa$ is the relativistic quantum number.

At this point, a comparison between our approach and the method
presented by GM can be made. There are two major differences between the approaches. The GM recipe used the analytical
radial wave function solutions of "Eq.~(\ref{dirac})" with $V(r)$
given by $-Ze^2/r$. The finite-size correction introduced by Rose
and Holmes~\cite{Ros51} has been later applied analytically to the obtained Fermi function. The advantage of numerically solving
"Eq.~(\ref{dirac})" was that we introduced the finite-size correction and the diffused nuclear surface correction by
choosing a proper electrostatic potential presented below. The
second difference between the two formalisms is the inclusion of the atomic screening correction. Whereas GM used the change of variables in the radial wave functions, we implemented this directly in the
electrostatic potential.

\begin{figure}[H]
\centering
{\includegraphics [width=3.5in]{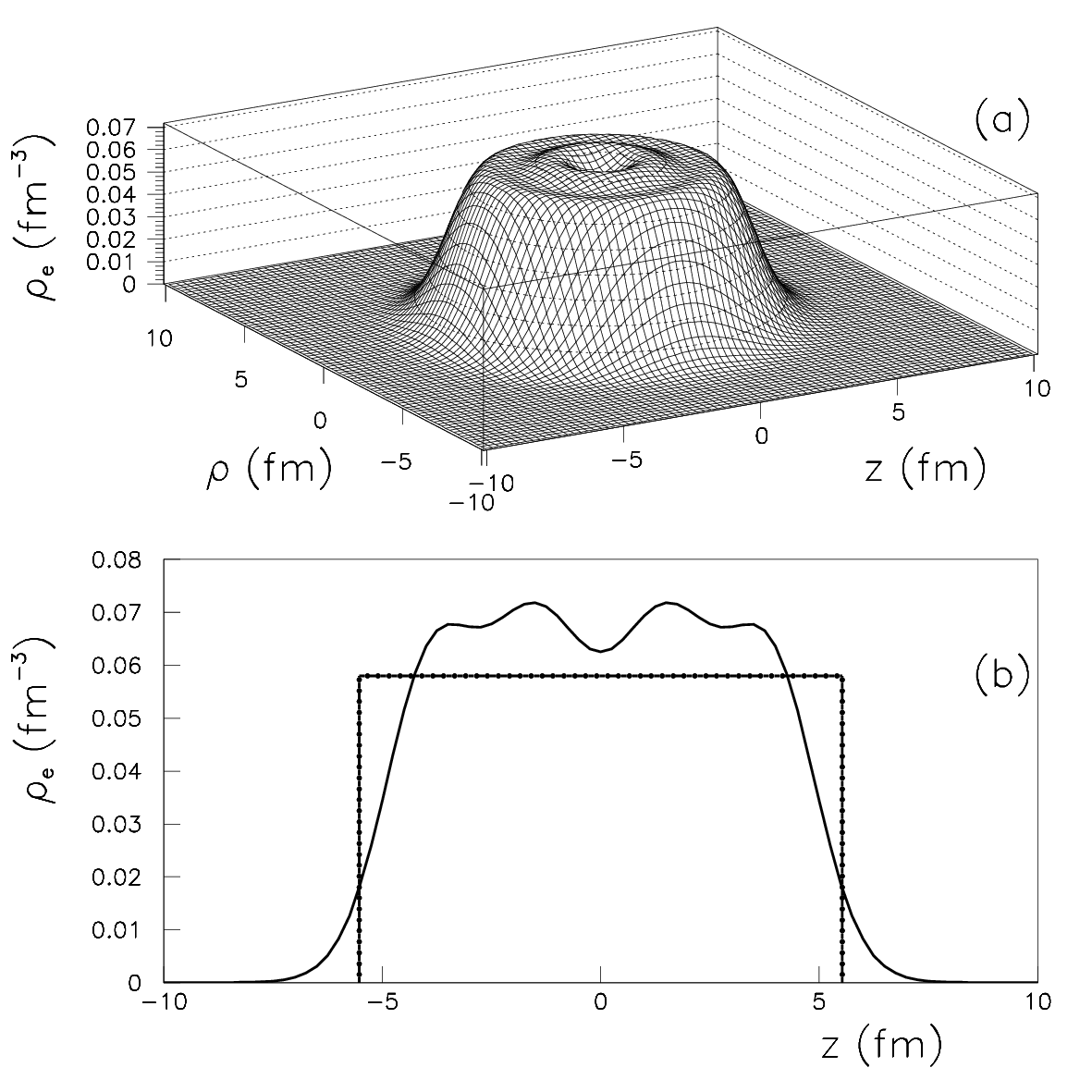}
   \centering \caption{\scriptsize (a) Realistic proton density for an atomic number  $Z$=40 represented in cylindrical coordinates. (b) Profile of the
 realistic proton density for $Z$ = 40 (thick line) compared with that given by a constant density approximation (dot-dashed line).  }
  \label{densfig}}
\end{figure}

The electrostatic potential used in this work, was adopted from
a more realistic proton density distribution in the nucleus, in comparison with a uniform one. The charge density
was introduced using
\begin{equation}\label{dens}
\rho_e(\vec{r})=\sum_{i}^{}(2j_i+1)v^2_i\left|\psi_i(\vec{r})\right|^2,
\end{equation}
where $\psi_i$ denotes proton wave function for the
spherical single particle state $i$ and $v_i$ stands for amplitude
of occupation. By solving Schr\"odinger equation using Woods-Saxon
potential the wave functions ${\psi_i}$, were obtained.
The (2$j_i$ + 1) term in "Eq.~(\ref{dens})" reflects the degeneracy
of spin. The difference between a uniform charge density and the one derived in this work can be found in "Fig.~(\ref{densfig}).

Coulomb potential was obtained by integrating the realistic proton
density distribution
\begin{equation}\label{potential}
V(Z,r)=\alpha\hbar c \int
\frac{\rho_e(\vec{r'})}{|\vec{r}-\vec{r'}|}d\vec{r'}.
\end{equation}

\begin{figure}[H]
\centering
	{\includegraphics [width=3.5in]{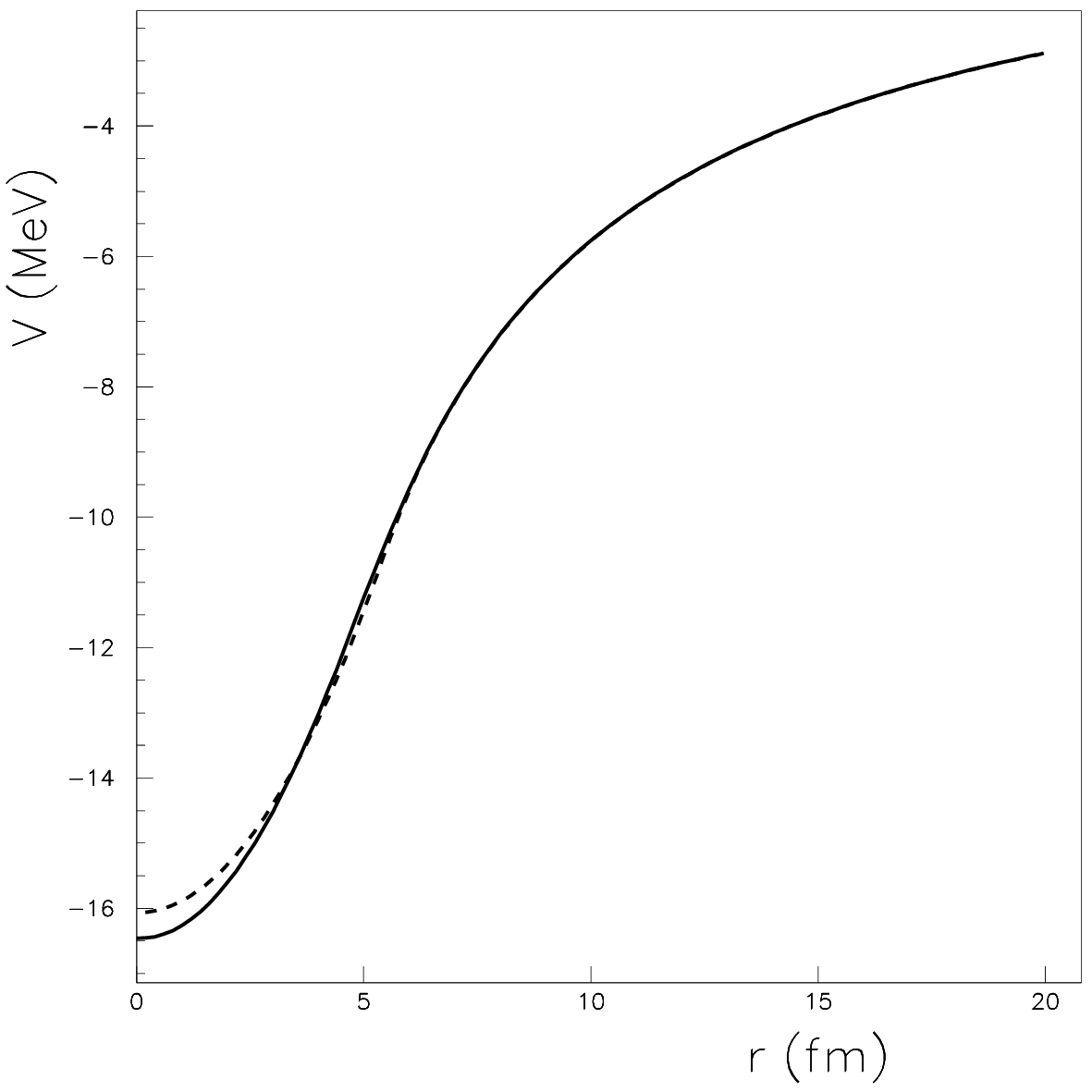}
		\centering \caption{\scriptsize Comparison between the Coulomb potential obtained
			with a realistic charge density displayed with a full
			line, and a potential obtained within a constant charge density plotted
			with a dashed line for a $Z$=40 residual nucleus.  }
		\label{coulomb}}
\end{figure}

The Coulomb potential obtained with the "Eq.~(\ref{potential})" is represented in
"Fig.~(\ref{coulomb})" with a full curve for a residual nucleus with $Z$=40 and $A$=90.
In the same figure the Coulomb potential obtained
with a constant charge density in the volume of the residual nucleus
is also displayed with a dashed line. For some nuclei, the differences
between the
two potentials can lead to spreads which amount 0.5\% in the Fermi function.
One considers that the potential obtained within the calculated proton
density is more realistic.

Solutions of "Eq.~(\ref{dirac})" with the electrostatic potential described by "Eq.~(\ref{potential})" include the finite-size and diffuse nuclear surface effects. Furthermore, we also
consider the atomic screening effect by changing the expression of
$V(r)$ with function $\phi(r)$, that is solution for Thomas-Fermi
equation

\begin{equation}\label{Thomas-Fermi}
\frac{d^2\phi}{dx^2}=\frac{\phi^{3/2}}{\sqrt{x}},
\end{equation}

with $x=r/b$, $b\approx0.8853r_bZ^{-1/3}$ and $r_b$ is the Bohr
radius. The solution $\phi(r)$ has been calculated using
Majorana recipe \cite{Esp02}. In case of $\beta^- / \beta^+$, the
effective potential $V_{\beta^{\mp}}$ was improved with the
help of screening function $\phi(r)$ as

\begin{equation}\label{Vmodified}
rV_{\beta^{\mp}}(Z,r)=(r V(Z,r)+ 1) \times \phi(r) - 1.
\end{equation}
$\beta$-decay half-lives were calculated by summing all
transition probabilities to daughter excited states lying within the Q$_{\beta}$ value 

\begin{equation}
T_{1/2} =( \sum_{0\leq E_{x}\leq Q_{\beta}} 1/t_{x})^{-1}
\label{hl},
\end{equation}
where t$_{x}$ are the partial half-lives:

\begin{equation}
t_{x} = \frac{C}{(g_{A}/g_{v})^{2}F_{A/V}B(E_{x})}
\label{phl}.
\end{equation}

In "Eq.~(\ref{phl})" C is a constant whose value was taken as
6143 s \cite{Har09}, for weak interaction $g_{A}$, $g_{v}$
denotes axial-vector and vector coupling constants,
respectively, having $g_{A}$/$g_{v}$= -1.2694 \cite{Nak10}, while
$E_x$ is the daughter excitation energy.
$F_{A/V}$ are the PSFs discussed above. B$(E_{x})$ are the reduced transition
probabilities for the Fermi and Gamow-Teller (GT) transitions. We
can express these reduced transition probabilities in form of NMEs
as follows

\begin{equation}
B_{F}(E_x) = \frac{1}{2I_{i}+1} \mid<x \parallel M_{F}
\parallel i> \mid ^{2}, \label{ftp}
\end{equation}

\begin{equation}
B_{GT}(E_{x}) =\frac{1}{2I_{i}+1} \mid<x\parallel M_{GT} \parallel
i>\mid^{2} \label{rtp}
\end{equation}

In "Eq.~(\ref{ftp})" and "Eq.~(\ref{rtp})", $I_{i}$
represents spin of parent state, $M_{F}$ and $M_{GT}$
denote the Fermi and GT transition operators,
respectively. Detailed description about the computation of
NMEs using pn-QRPA theory could be seen in Refs.
\cite{Hir93,Sta90}.

We further explored the effect of pairing gaps on calculated PSFs and  $\beta$-decay
half-lives. Pairing gap computation proved crucial for the current calculation. For the calculation of pairing gap, in units of MeV, we employed the so called 3-point formulae. These formulae are a function of neutron and proton separation energies
as shown in Eqs.~(\ref{PP}~-~\ref{NN})
\begin{equation}\label{PP}
\Delta_{pp}=
\frac{1}{4}(-1)^{Z+1}[S_{p}(A+1,Z+1)-2S_{p}(A,Z)+S_{p}(A-1,Z-1)]
\end{equation}
\begin{equation}\label{NN}
\Delta_{nn}=
\frac{1}{4}(-1)^{A-Z+1}[S_{n}(A+1,Z)-2S_{n}(A,Z)+S_{n}(A-1,Z)]
\end{equation}

For further details of the calculation of PSFs and
NMEs, we refer to \cite{Sab16, Ish19}. \\ \\

\section{Results}

Table 1 shows the PSFs computed with the current and GM recipes for the six even-even selected nuclei. $Q$-values were taken from Ref. \cite{Aud17}. The two results differ, at the maximum by 10 percent.

A comparison between measured and the calculated values of $\beta$-decay half-lives for the six selected nuclei are
shown in Table 2. Measured half-lives appearing in the second
column were taken from \cite{Aud17}. Computed half-lives in the third
column were obtained using the traditional PSF recipe of GM and associated NMEs
from the pn-QRPA model. The last column presents
the calculated half-lives employing the newly introduced PSFs and labeled $(C)$ (current). The associated NMEs were calculated within the framework of pn-QRPA model. Half-life values in Table 2 are stated in units of seconds. It is noted that the calculated half-lives using the new prescription of PSFs are in good agreement with the experimental data.

Table 3 presents the state-by-state computation of partial half-lives for $^{20}$O (the largest difference between the two calculations as noted in Table~2) and $^{34}$Si (the smallest difference between the two calculations). The values of pn-QRPA model calculated daughter excitation energies ($E_x$),  $Q_{\beta^-}$ (= $m_p- m_d - E_x$), PSFs, NMEs, and branching ratios
I$_{(\beta^-)}$ are also given in Table 3. For the case
of $^{20}$O, we note that the partial half-lives and branching ratios
to the last four levels contribute to the big difference between
the two calculations. The computed branching
ratios and partial half-lives are in decent agreement for all states in case of   $^{34}$Si.
\section{Conclusions}
The newly introduced prescription of PSFs were applied to compute $\beta$-decay half-lives of even-even nuclei. The required NMEs were
computed within the framework of pn-QRPA model. A
small difference between the traditionally and newly introduced
recipe of PSFs was noted. A similar difference is also reported
for the computed values of $\beta$-decay half-lives using the two recipes of PSFs. The newly introduced recipe of PSFs resulted in calculated $\beta$-decay half-lives in better agreement with measured ones.


\begin{table}[h!]
  \caption{\label{BMPSF} Calculated total phase space factors $F_{\beta^{-}}$ for $\beta^{-}$ decay, with  Q-values from \cite{Aud17}.}
    {\begin{tabular}{@{}ccccccc@{}} \toprule
Nucleus & $Q_{\beta^-}$ (MeV) \cite{Aud17}  & $F^{(GM)}_{\beta^{-}}$\cite{Gov71} & $F^{(C)}_{\beta^{-}}$ \\
            \hline $^{20}$O&    3.81366 & 5864.37  & 5849.69 \\
            $^{24}$Ne&   2.46630 & 70.6367   & 70.4907 \\
            $^{34}$Si&   4.59170 & 5531.61   & 6167.34 \\
            $^{54}$Ti&   4.27300 & 26816.4   & 26556.5  \\
            $^{62}$Fe&   2.54600 & 2452.09   & 2422.20  \\
            $^{98}$Zr&   2.23800 & 3372.44   & 3294.33  \\
            \hline
   \end{tabular} \label{BMPSF}}
\end{table}

 \begin{table}[h!]
 \caption{\label{BMHL} Comparison of measured \cite{Aud17} and computed half-lives using PSFs by GM prescription of \cite{Gov71} and
 by newly introduced recipe (C) \cite{Sab16}. Half-life values are stated in units of $s$.}
    {\begin{tabular}{@{}ccccccc@{}} \toprule
        Nucleus & $T_{1/2}^{(EXP)}$ (s) \cite{Aud17} & $T_{1/2}^{(GM)}$ (s) \cite{Gov71}& $T_{1/2}^{(C)}$ (s)\\
        \hline $^{20}$O      & (1.351$\pm$0.005) E+01 &2.12E+01         &1.49E+01  \\
        $^{24}$Ne            & (2.028$\pm$0.012) E+02 &2.26E+02         &2.05E+02  \\
        $^{34}$Si            & (2.770$\pm$0.200) E+00 &3.01E+00         &2.87E+00  \\
        $^{54}$Ti            & (2.100$\pm$1.000) E+00 &3.14E+00         &2.41E+00  \\
        $^{62}$Fe            & (6.800$\pm$0.200) E+01 &9.24E+01         &7.68E+01  \\
        $^{98}$Zr            & (3.070$\pm$0.040) E+01 &4.06E+01         &3.33E+01  \\
    \hline
\end{tabular} \label{BMHL}}
\end{table}

\begin{table}[h!]
\caption{\label{BMPHLS} State-by-state comparison of computed $\beta$-decay half-lives using the GM and newly introduced recipes for computation of PSFs. The daughter energy levels, NMEs, $Q$ values, partial half-lives
    t$_{f}$  and branching ratio I$_{(\beta^-)}$ for $\beta^{-}$-decay to calculated daughter states are also shown.}
    {\begin{tabular}{@{}ccccccccc@{}} \toprule
$^{20}$O\\
        \hline E$_{x}$(MeV) & NME & $Q_{\beta^-}$ (MeV) & $F^{(GM)}_{\beta^{-}}$\cite{Gov71} & $F^{(C)}_{\beta^{-}}$ & t$_{f}^{(GM)}$\cite{Gov71} & t$_{f}^{(C)}$ &  I$^{(GM)}_{(\beta^-)}$ \cite{Gov71} & I$^{(C)}_{(\beta^-)}$\\
        \hline         0.00400 & 0.00000    & 3.80967 & 1689.06 & 1684.56& 1.40644E+06& 4.87374E+05& 0.0020&0.0030\\
                       0.24900 & 0.01929    & 3.56411 & 1252.16 & 1249.04& 1.65717E+02& 4.16003E+01& 12.810&17.260\\
                       0.26200 & 0.03908    & 3.55175 & 1232.83 & 1229.70& 8.30920E+01& 4.80330E+01& 25.549&35.898\\
                       0.54200 & 0.09132 & 3.27209 & 854.878 & 852.899 & 5.12796E+01& 9.48294E+01& 41.399&31.091\\
                      0.55800 & 0.04569 & 3.25521 & 835.436 & 833.484 & 1.04884E+02& 6.66259E+03& 20.240&15.748\\
        \hline$^{34}$Si\\ \hline0.70800  &0.26574  &  3.88335& 2167.55 & 2414.32 & 6.94985E+00& 6.65015E+00& 43.421&43.163\\
                       0.84300 & 0.00000    & 3.74878 & 1850.47 & 2062.28 & 1.54665E+08& 1.60074E+08& 0.0000&0.0000\\
                       1.07200 & 0.53771    & 3.51923 & 1395.43 & 1557.01 & 5.33510E+00& 5.05171E+00& 56.563&56.821\\
                       2.66400 & 0.00000    & 1.92758 & 102.739 & 115.983 & 2.09311E+09& 1.19578E+09& 0.0000&0.0000\\
                       3.47600 & 0.01324    & 1.11531 & 11.0202 & 12.6278 & 2.74395E+04& 2.64834E+04& 0.0110&0.0110\\
                       3.82500 & 0.00000    & 0.76635 & 2.57740 & 2.99066 & 4.10905E+11& 3.83127E+11& 0.0000&0.0000\\
                       3.96800 & 0.05621    & 0.62359 & 1.19018 & 1.39208 & 5.98407E+04& 5.67186E+04& 0.0050&0.0050\\
                       4.06900 & 0.00000    & 0.52264 & 0.62227 & 0.73302 & 3.54261E+12& 3.28252E+12& 0.0000&0.0000\\
                       \hline
   \end{tabular} \label{BMPHL}}
\end{table}
\acknowledgments{J.-U. Nabi would like to acknowledge the support of the Higher
Education Commission Pakistan
through project numbers 5557/KPK
/NRPU/R$\&$D/HEC/2016, 9-5(Ph-1-MG-7)/PAK-TURK
/R$\&$D/HEC/2017 and Pakistan Science Foundation through project
number PSF-TUBITAK/KP-GIKI (02).\\ S. Stoica would like to acknowlege the support of MCI through project number PN19-030102-INCDFM.}


\reftitle{References}



\begin{thebibliography}{999}
\bibitem[Weinberg (2009)]{Wei09}
S. Weinberg. V-A was The Key. {\em J. Phys. Conf. Ser.} {\bf 2009}, {\em 196}, 012002.
\bibitem[Moller (2003)]{Mol03}
P. M\"oller, B. Pfeiffer and K. L. Kratz. New calculations of gross $\beta$-decay properties for astrophysical applications: Speeding-up the classical r process. {\em Phys. Rev. C.} {\bf 2003}, {\em 67}, 055802.
\bibitem[Ni (2014)]{Ni14}
D. Ni and Z. Ren. $\beta$-decay rates of r-process waiting-point nuclei in the extended quasiparticle random-phase approximation {\em J. Phys. G: Nucl. Part. Phys} {2014}, {\em 41}, 025107.
\bibitem[Ren (2014)]{Ren14}
Y. Ren and Z. Ren. Systematic law for half-lives of double-$\beta$ decay with two neutrinos. {\em Phys. Rev. C}, {2014}, {\em 89}, 064603.
\bibitem[Groote (1976)] {Gro76}
H. V. Groote, E. R. Hilf, K. Takahashi. A new semiempirical shell correction to the droplet model: Gross theory of nuclear magics. {\em At. Data Nucl. Data Tables}, {1976}, {\em 17}, 418-427.
\bibitem[Halpern (1970)] {Hal70}
T. A. Halpern. Static Coulomb Correction to Beta Decay Arising from the Nuclear Charge Change: A Nonperturbative Approach. {\em Phys. Rev. C } {1970} {\em 1}, 1928-1938.
\bibitem[Roman (1965)] {Rom65}
P. Roman. Advanced Quantum Theory. Addison-Wesley Publishing Company, Inc., Reading, Massachusetts, 1965.
\bibitem[Behrens (1968)] {Beh68}
H. Behrens and W. B\"{u}hring. ft values of superallowed 0-0 transitions {\em Nucl. Phys. A } {1968} {\em 106}, 433.
\bibitem[Behrens (1970)] {Beh70}
H. Behrens and W. B\"{u}hring. On the sensitivity of $\beta$-transitions to the shape of the nuclear charge distribution. {\em Nucl. Phys. A } {1970} {\em 150}, 481.
\bibitem[Konopinski (1941)] {Kon41}
E. J. Konopinski and G. E. Uhlenbeck. On the Fermi Theory of $\beta$-Radioactivity. II. The "Forbidden" Spectra. {\em Nucl. Phys. Rev. } {1941} {\em 60}, 308.
\bibitem[Hayen (2018)] {Hay18}
L. Hayen, N. Severijns, K. Bodek, D. Rozpedzik and X. Mougeot. High precision analytical description of the allowed $\beta$ spectrum shape. {\em Rev. Mod. Phys. } {2018} {\em 90}, 015008.
\bibitem[Fermi (1934)] {Fer34}
E. Fermi. Versuch einer Theorie der $\beta$-Strahlen. I. {\em Zeitschrift fur Phys. } {1934} {\em 88}, 161.
\bibitem[Wilkinson (1990)] {Wil90}
D. H. Wilkinson. Evaluation of $\beta$-decay: II. Finite mass and size effects {\em Instruments Methods Phys. Res. A. } {1990} {\em 290}, 509.
\bibitem[Salvat (1991)] {Sal91}
F. Salvat and R. Mayol. Accurate numerical solution of the Schrödinger and Dirac wave equations for central fields. {\em Comp. Phys. Commun.} {1991} {\em 62}, 65.
\bibitem[Salvat (1995)] {Sal95}
F. Salvat, J. M. Fernandez-Varea and W. Jr. Williamson. Accurate numerical solution of the radial Schrödinger and Dirac wave equations. {\em Comp. Phys. Commun.} {1995} {\em 90}, 151.
\bibitem[Sabin (2016)] {Sab16}
S. Stoica, M. Mirea, O. Ni\c{t}escu, J. -U. Nabi  and M. Ishfaq. New Phase Space Calculations for $\beta$-Decay Half-Lives {\em Adv. High Energy Phys} {2016} {\em 2016}, ID 8729893.
\bibitem[Gove (1971)] {Gov71}
N. B. Gove and M. J. Martin. Log-f tables for beta decay. {\em Nucl. Data. Tables.} {1971} {\em 10}, 205.
\bibitem[Behrens (1969)] {Beh69}
H. Behrens, J. J\"{a}necke. Numerical Tables for Beta-Decay and Electron Capture. {1969}.
\bibitem[Rose (1961)] {Ros61}
M. E. Rose. Relativistic Electron Theory {\em Wesley-VCH.} {1961}.
\bibitem[Rose (1951)] {Ros51}
M. E. Rose and D. K. Holmes. Oak Ridge National Laboratory Report ORNL-1022. {\em Phys. Rev} {1951} {\em 83}, 190.
\bibitem[Esposito (2002)] {Esp02}
S. Esposito. Majorana solution of the Thomas–Fermi equation. {\em Am. J. Phys} {2002} {\em 70}, 852.
\bibitem[Hardy (2009)] {Har09}
J. C. Hardy and I. S. Towner. Superallowed $0^{+}$ $0^{+}$ nuclear $\beta$ decays: A new survey with precision tests of the conserved vector current hypothesis and the standard model. {\em Phys. Rev. C.}, {2009}, {\em 79}, 055502.
\bibitem[Nakamura (2010)] {Nak10}
K. Nakamura. Review of particle physics. {\em J. Phys. G: Nucl. Part. Phys} {2010} {\em 37}, 075021.
\bibitem[Hirsh (1993)] {Hir93}
M. Hirsch, A. Staudt, K. Muto, and H. V. Klapdorkleingrothaus. Microscopic Predictions of $\beta^{+}$/EC-Decay Half-Lives {\em At. Data and Nucl. Data Tables} {1993} {\em 53}, 165-193.
\bibitem[Staudt (1990)] {Sta90}
A. Staudt, E. Bender, K. Muto and H. V. Klapdor- Kleingrothaus. Second generation microscopic predictions of beta decay half-lives of neutron rich nuclei. {\em At. Data. Nucl. Data. Tables} {1990} {\em 44}, 79.
\bibitem[Ishfaq (2019)] {Ish19}
M. Ishfaq, J. -U. Nabi, O. Ni\c{t}escu, M. Mirea and S. Stoica. Study of the Effect of Newly Calculated Phase Space Factor on
$\beta$-Decay Half-Lives {\em Adv. High Energy Phys} {2019} {\em 2019}, ID 5783618.
\bibitem[Audi (2017)] {Aud17}
G. Audi, F. G. Kondev, M. Wang, W. J. Huang and S. Naimi. The NUBASE2016 evaluation of nuclear properties. {\em Chin. Phys. C.} {2017} {\em 41}, 030001.
\end{thebibliography}



\end{document}